\documentclass[11pt,a4paper]{article}
\pdfoutput=1
\usepackage{jcappub}
\usepackage{float}
\usepackage{subfigure}
\title{3.55 keV photon lines from axion to photon conversion in the Milky Way and M31}
\author[a]{Joseph P. Conlon,}
\author[a]{Francesca V. Day}
\affiliation[a]{Rudolf Peierls Centre for Theoretical Physics, University of Oxford,\\
1 Keble Road, Oxford, OX1 3NP, United Kingdom}
\emailAdd{j.conlon1@physics.ox.ac.uk}
\emailAdd{francesca.day@physics.ox.ac.uk}

\abstract{

We further explore a scenario in which the recently observed 3.55 keV photon line arises from dark matter decay to an axion-like particle (ALP) of energy 3.55 keV,
which then converts to a photon in astrophysical magnetic fields. This ALP scenario is well-motivated by the observed morphology of the 3.55 keV flux.
For this scenario we study the expected flux from dark matter decay in the galactic halos of both the Milky Way and Andromeda (M31). The Milky Way magnetic field is asymmetric about the galactic centre, and so the resulting 3.55 keV flux morphology differs significantly from the case of direct dark matter decay to photons. However the Milky Way magnetic field is not large enough to generate an observable signal, even with ASTRO-H. In contrast, M31 has optimal conditions for $a \to \gamma$ conversion and the intrinsic signal from M31 becomes two orders of magnitude larger than for the Milky Way,
comparable to that from clusters and consistent with observations.
}

\arxivnumber{}

\newcommand{\be}{\begin{equation}}
\newcommand{\ee}{\end{equation}}
\newcommand{\bea}{\begin{eqnarray}}
\newcommand{\eea}{\end{eqnarray}}

\newcommand{\ti}{\times}

\begin{document}

\maketitle
\flushbottom

\section {Introduction}

Recent results suggest the existence of a 3.55 keV photon line in a number of galaxy clusters and in the Andromeda galaxy (M31) \cite{Bulbul, Boyarsky}. As this line has been observed with both XMM-Newton PN and MOS cameras, and also with both ACIS-I and ACIS-S configurations on the Chandra satellite, it does not appear to be an instrumental effect. While the observations of clusters in \cite{Bulbul} leave open the possibility that this is an atomic line from the hot ionised intracluster medium (ICM), the simultaneous observation of the line in M31 in \cite{Boyarsky} argues against this interpretation: unlike clusters, galaxies are not suffused with hot multi-keV gas.\footnote{ Since this paper was written, the paper \cite{JeltemaProfumo} claims that there is no significant line emission in M31, while in response \cite{BoyarskyII} argues that this analysis is incorrect and the original claim of a 3.5 keV line in M31 still stands. It is beyond the scope of this paper to arbitrate between these competing claims.}
 While due scepticism is required, the most exciting interpretation of this as yet unidentified line is as a result of dark matter decay or annihilation to photons.

We focus in this paper on the possibility considered in \cite{ALP355} that the line arises from dark matter decay to an  axion like particle (ALP), which subsequently converts to a photon in astrophysical magnetic fields \cite{ALP355}.\footnote{Other papers discussing the 3.55 keV line signal include \cite{1,2,3,4,5,6,7,8,9,10,11,12,13,14,15,16,17,18,19,20,21,22,23,24,25,26,27}.} This process of ${\rm DM} \to a \to \gamma$ can explain morphological features of the signal of \cite{Bulbul} that are inconsistent with the direct decay of dark matter to photons. For example, in \cite{Bulbul} the signal from Perseus was found to be much stronger than for the stacked sample of distant clusters, and within Perseus both XMM-Newton and Chandra data show that the signal peaks sharply in the central cool core region of the cluster, whereas the dark matter distribution is much broader.

The underlying physical reason for this is that an $a \to \gamma$ conversion probability is proportional to the square of the magnetic field transverse to the direction of travel, and is thus enhanced in regions of high magnetic field. While Faraday Rotation Measure (RM) studies show that clusters generally have magnetic fields of $O \left( \mu {\rm G} \right)$ (for a review see \cite{Feretti}), this can vary by a factor of a few both between clusters and within a cluster. RM studies also show significantly enhanced magnetic fields in the central regions of cool core clusters such as Perseus. This scenario therefore predicts an enhanced signal in the centre of cool core clusters, consistent with the current data. Further details of the general predictions of this scenario are described in \cite{ALP355}.

If the 3.55 keV line does indeed arise from the process ${\rm DM} \to a \to \gamma$, we also expect a line signal from dark matter decays in galactic halos, followed by conversion in the galactic magnetic field.
Various aspects of ALP to photon conversion in the Milky Way's magnetic field have
been previously considered in \cite{brockway,Simet,12070776,13021208,Wouters,Fairbairn}.

The main purpose of this article is to calculate the strength and morphology of the resulting 3.55 keV photon line from the Milky Way's dark matter halo decay, and to examine detection possibilities. As the 3.55 keV line has also been observed in M31, we also discuss the M31 magnetic field and the expected strength of the resulting signal, extending the very brief discussion in \cite{ALP355}.

The paper is organised as follows.
In sections 2 and 3 we summarise the relevant properties of ALPs and the Milky Way. In section 4 we describe our results for both the Milky Way and M31,
 and in section 5 we conclude.

\section {ALPs}

The existence of ALPs is theoretically well motivated, as they arise generically in string theory compactifications.
The properties and phenomenology of ALPs are reviewed in \cite{Ringwald}.
The basic ALP Lagrangian is
\begin{equation}
\mathcal{L} =  \frac{1}{2} \partial_{\mu} a \partial^{\mu} a - \frac{1}{2} m_a^2 a^2 + \frac{a}{M} {\bf E} \cdot {\bf B},
\end{equation}
but throughout this paper we assume a massless ALP with $m_{a} = 0$.  The coupling $M$ is key to the phenomenology and determines the strength of interactions between ALPs and electromagnetism.


The coupling $a {\bf E} \cdot {\bf B}$ implies that in the presence of a background magnetic field, the ALP $a$ mixes with the photon in a manner analogous to neutrino oscillations. The formalism of these $a \leftrightarrow \gamma$ oscillations is developed in full in \cite{Sikivie, Raffelt}. The equation of motion for the ALP-photon vector is
\begin{equation}
\label{propfree}
\left( \omega + \left( \begin{array}{ccc}
\Delta_{\gamma} & \Delta_{F} & \Delta_{\gamma a x} \\
\Delta_{F} & \Delta_{\gamma} & \Delta_{\gamma a y} \\
\Delta_{\gamma a x}  & \Delta_{\gamma a y} & \Delta_{a} \end{array} \right)
 - i\partial_{z} \right) \left( \begin{array}{c}
\mid \gamma_{x} \rangle\\
\mid \gamma_{y} \rangle\\
\mid a \rangle \end{array} \right) = 0,
\end{equation}
where $\Delta_{\gamma} = \frac{-\omega_{pl}^{2}} {2 \omega}$, the plasma frquency is $\omega_{pl} = \left( 4 \pi \alpha \frac{n_{e}}{m_{e}} \right) ^ {\frac{1}{2}}$,
and $\Delta_{a} = \frac{-m_{a}^{2}}{\omega}$. The mixing is controlled by $\Delta_{\gamma a i} = \frac{B_{i}}{2M}$. Here, $x$ and $y$ refer to the directions transverse to the direction of travel. $\Delta_{F}$ parameterises Faraday rotation between the two photon polarizations. As this effect is not important for ALP to photon conversion, which
is concerned with the transition from the ALP state to either of the photon polarizations, 
 we neglect it in our subsequent discussion.

As the mixing term giving the quantum amplitude to convert from an $a$ state to a photon state is proportional to $\frac{B_{\perp}} {M}$,
it is clear that the conversion probability $P$ is proportional to $B_{\perp}^{2} = B_{x}^{2} + B_{y}^{2}$ and to $\frac{1}{M^{2}}$. The solutions of the equation of motion in various limiting cases are discussed further in \cite{Coma, ALP355}. While a general magnetic field requires numerical solution, much of the physics can be understood from the single domain case. Here we can identify two angles associated with the propagation, $\theta$ and $\Delta$. Their analytic expressions for $m_a = 0$ are
\bea
\tan \left( 2 \theta \right) & = & \frac{ 2 B_{\perp} \omega}{M \omega_{pl}^2}, \\
\Delta & = & \frac{\omega_{pl}^2 L}{4 \omega}.
\eea
which translate into
\bea
{\rm tan} \left( 2 \theta \right) & = & 10.0 \times 10^{-3} \times \left( \frac{10^{-3} \, {\rm cm}^{-3}}{n_{e}} \right) \left( \frac{B_{\perp}}{1 \, \mu {\rm G}} \right) \left( \frac{\omega}{3.55 \, {\rm keV}} \right) \left( \frac{10^{13} \, {\rm GeV}}{M} \right), \\
\Delta & = & 0.015 \times \left( \frac{n_{e}}{10^{-3} \, {\rm cm}^{-3}} \right) \left(\frac{3.55 \, {\rm keV}}{\omega} \right) \left( \frac{L}{1 \, {\rm kpc}} \right).
\eea
For a single domain of length $L$, the conversion probability is then
\be
P(a\ \to \gamma) = \sin^2 \left( 2 \theta \right) \sin^2 \left( \frac{\Delta}{\cos 2 \theta} \right).
\ee
For the case $\theta, \Delta \ll 1$, and propagation through a total distance of $R$ consisting of $R/L \gg 1$ separate domains between
which the magnetic field is randomised, we may use the illustrative small angle approximation \cite{ALP355}:
\begin{equation}
P \simeq 2.3 \times 10^{-8} \left( \frac{L}{1 \, {\rm kpc}} \frac{R}{1 \, {\rm kpc}} \right) \left( \frac{B_{\perp}}{1 \, \mu {\rm G}} \frac{10^{13} \, {\rm GeV}}{M} \right)^{2},
\end{equation}
where $R$ is the typical size of the galaxy or cluster and we assume $R/L$ domains of size $L$. For a 3.55 keV ALP propagating through the Milky Way, we have always $\theta \ll 1$ and mostly but not always $\Delta \ll 1$. Within the galactic plane, both $n_e$ and the coherence length of the regular field
are maximised, and we do not have $\Delta \ll 1$.

The model we analyse here assumes that dark matter decays to produce a monoenergetic ALP of energy 3.55 keV \cite{ALP355}. This can arise either from a decay 
$\psi \to \chi a$, where $\psi$ is the parent dark matter particle, $\chi$ is a massless daughter and $a$ is the axion-like particle, or from a decay to $\psi \to a a$, which produces
two ALPs.
The expected line signal from ALPs is set by the product $\tau M^{2}$, where $\tau$ is the dark matter to ALP decay time.\footnote{Note that $\tau$ is not necessarily the total dark matter lifetime, as there may be additional hidden sector decays.} In \cite{ALP355}, it was shown that reproducing the observed signal strength of \cite{Bulbul} from the
 sample of stacked galaxy clusters requires:
\begin{equation}
\label{tauLife}
\tau \sim 5 \times 10^{24} \, {\rm s} \left(\frac{10^{13} \, {\rm GeV}}{M} \right) ^{2}.
\end{equation}
This assumes that the ALP is produced by a decay $\psi \to \chi a$. The alternative process $\psi \to a a$ simply requires a change in decay time by a factor of two.
We also assume that for a typical galaxy cluster and  with $E_a = 3.55 \, {\rm keV}$, $P(a \to \gamma) = 10^{-3} \left( \frac{10^{13} \, {\rm GeV}}{M} \right)^2$. These numerical values are extracted from propagation of ALPs through a detailed simulation of the magnetic field for the central $1 \, \hbox{Mpc}^3$ of the Coma cluster \cite{Coma}, extending the estimates of \cite{13053603}. While the conversion probability for the central region of the Coma cluster can clearly only be an approximation for the stacked sample of cluster in the analysis of \cite{Bulbul}, we use it as an approximate guide to the required dark matter lifetime.

We will set $\tau M^2$ from equation (\ref{tauLife}) when calculating the signal strength expected from the Milky Way and M31. However, we re-emphasise that the value in
equation (\ref{tauLife}) is based on assuming that the conversion probabilities calculated in \cite{Coma} for the central region of the Coma cluster is on some level typical of clusters in general. This may in fact be either an over- or under- estimate of the actual  conversion probabilities averaged over the stacked cluster sample of \cite{Bulbul}.
We note in addition there is a potential systematic error in case the use of RMs to measure cluster magnetic fields is systematically inaccurate.

While the morphological distribution we calculate in section 4 below will be robust against variations in $\tau$, the overall line strength we find will
inherit this uncertainty. However as the 3.55 keV line signal has also been observed in M31 \cite{Boyarsky}, this bounds this uncertainty to a factor of a few (unless there are additional
large systematic errors in the magnetic field determinations in M31).


The propagation equation (\ref{propfree}) applies for a pure electron plasma. However, the propagation of photons in the Milky Way can be
significantly affected by photoelectric absorption. Following \cite{EBL}, we model this absorption by use of a density matrix formulation (the formalism for the evolution of open quantum systems is developed more formally in \cite{Bertlmann}):
\begin{equation}
H = \left( \begin{array}{ccc}
\Delta_{\gamma} & \Delta_{F} & \Delta_{\gamma a x} \\
\Delta_{F} & \Delta_{\gamma} & \Delta_{\gamma a y} \\
\Delta_{\gamma a x}  & \Delta_{\gamma a y} & \Delta_{a} \end{array} \right)
 -
 \left( \begin{array}{ccc}
i \frac{\Gamma}{2} & 0 & 0 \\
0 & i \frac{\Gamma}{2} & 0 \\
0  & 0 & 0 \end{array} \right) = M - iD ,
\end{equation}
\begin{equation}
\rho =  \left( \begin{array}{c}
\mid \gamma_{x} \rangle\\
\mid \gamma_{y} \rangle\\
\mid a \rangle \end{array} \right) \otimes
\left( \begin{array}{ccc}
\mid \gamma_{x} \rangle & \mid \gamma_{y} \rangle & \mid a \rangle \end{array} \right) ^{*}  \\ ,
\end{equation}
\begin{equation}
\rho(z) = e^{-iHz} \rho(0) e^{iH^{\dagger}z} .
\end{equation}
The damping parameter is $\Gamma = \sigma_{{\rm eff}} \left( n_{HI} + 2n_{H2} \right)$, where $n_{HI} + 2n_{H_2}$ is the density of neutral hydrogen.
 Photoelectric absorption is conventionally parametrised by the neutral hydrogen density, although at $E \sim 3.55 \, \hbox{keV}$ hydrogen itself is not so important and
 it is heavier elements that play a dominant role. We also note that the standard values for $\sigma_{eff}$ assume solar abundances for metals. However, as
 we will find that absorption only has a relatively small effect on the signal, we can neglect possible changes in abundances away from solar values.

In practice, we discretize the field into domains of length $\delta z$ so that:
\begin{equation}
\rho_{k} = e^{-iH_{k} \delta z} \rho_{k - 1} e^{iH^{\dagger}_{k} \delta z},
\end{equation}
where $\rho_{k}$ is the density matrix in the $k$th domain and $H_{k}$ is the Hamiltonian defined using the magnetic field, free electron density and hydrogen density in the centre of the $k$th domain.


\section {The Milky Way Dark Matter Halo and Magnetic Field}

 To determine the strength and morphology of the photon line, we require models for the Milky Way dark matter halo, magnetic field, electron density and neutral hydrogen abundance. For the dark matter profiles, we calculate the expected flux based on both NFW and Einasto profiles, using parameters from \cite{Fermi}. The NFW profile is \cite{NFW}:
\begin{equation}
\rho_{NFW}\left( r \right) = \frac{\rho_{s}}{\left( \frac{r}{r_{s}} \right) \left( 1 + \frac{r}{r_{s}} \right)^{2}},
\end{equation}
where $r$ is the spherical radius from the galactic centre, $r_{s} = 20 \, {\rm kpc}$ and $\rho_{s}$ is implicitly set by requiring the local dark matter density $\rho_{NFW}\left( R_{\odot} \right) = 0.4 \, {\rm GeV cm}^{-3}$, with $R_{\odot} = 8.5 \, \hbox{kpc}$.

The Einasto dark matter profile is
\begin{equation}
\rho_{E}\left( r \right) = \rho_{s} {\rm exp} \left[ \frac{-2}{\alpha} \left( \left(\frac{r}{r_{s}} \right)^{\alpha} - 1 \right) \right],
\end{equation}
with $\alpha = 0.17$, $r_{s} = 20 \, {\rm kpc}$ and $\rho_{s}$ is again set by $\rho_{E}\left( R_{\odot} \right) = 0.4 \, {\rm GeV cm}^{-3}$.

For the Milky Way magnetic field, we use the recent model of Jansson and Farrar, given in \cite{Farrar} and illustrated in Figure 1, based on 40,000 extragalactic Faraday RMs. The details and numerical parameters of the model are found in \cite{Farrar} and here we simply summarise the field. This model contains both a regular part and a random/striated component. The regular part has structure on the scale of the Milky Way and contains three elements: a disk field, a toroidal halo element and an out-of-plane X-component.  The disk field is directed in the plane of the galaxy and follows the structure of the spiral arms, although with some field reversals between different spirals. The halo field is also purely azimuthal, but extends out of the galaxy and with a greater radial extent to the South of the galaxy than to the North. The X-field component points out of the galactic plane.

We note here that this field model artificially excludes a 1 kpc sphere at the centre of the galaxy ($B$ is set to zero in this region). This is due to the difficulties in determining the magnetic field in this region. However, on physical grounds it is in this region we expect the magnetic field strength to be largest. We therefore expect this model to underestimate the conversion probabilities for ALPs passing directly through the galactic centre.

The random/striated component contains a striated field, which is locally aligned with the regular field but of random sign, and a random field, which is unaligned with the regular component.  These components are expected to be inserted into the galaxy through supernova outflows. Although the striated and random field components are actually larger than the coherent field, their coherence length is much less, of order 100 pc. This implies their contribution to $P(a \to \gamma)$ is much smaller than that of the coherent field, and so we can self-consistently neglect these in our calculations.

Let us now justify our neglect of the random and striated fields in our calculations. As above, the coherence length of the random and striated fields are expected to be $L \lesssim 100  \, \hbox{pc}$, and the typical value of the random magnetic field in the disk is $\langle B \rangle \sim 3 \, \mu G$. Using the small angle approximation described in Section 2 (note that the low value of $L$ here means we have $\Delta \ll 1$ for most of the galaxy) with $L \sim 100 \, {\rm pc}$, $R \sim 30 \, {\rm kpc}$, $B_{\perp} \sim 3 \, \mu {\rm G}$ and $M = 10^{13} \, {\rm GeV}$, we find the conversion probability due to the random field alone is $P_{rand} \sim 10^{-7}$. This is smaller by an order of magnitude than the conversion probabilities arising from the regular field for a 3.55 keV ALP with $M = 10^{13} \, {\rm GeV}$ created at $R \sim 30 \, {\rm kpc}$ from the Earth.

We therefore find that even though the random component of the magnetic field is significantly larger than the regular component, its much shorter coherence length means that it can be self-consistently neglected when considering ALP-to-photon conversion.

\begin{figure} [h]
\includegraphics[scale=0.7]{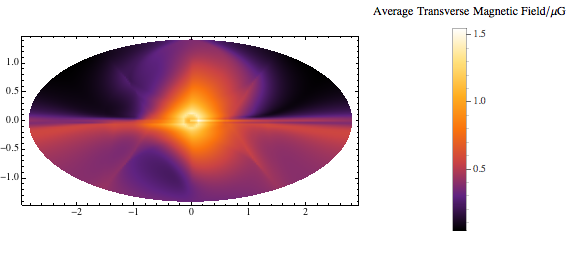}
\caption{The average regular transverse magnetic field experienced by an ALP on a path starting 20 kpc from the Earth and ending at the Earth. Galactic longitude increases to the right and galactic latitude increases vertically. The centre of the plot corresponds to the direction of the galactic centre.}
\end{figure}

We use a simple model for the electron density $n_{e}$, following \cite{Gomez}, comprising thick disk and thin disk components with
vertical scale heights 1.10 kpc and 0.04 kpc respectively.\footnote{Strictly this is not entirely self-consistent, as the Jansson and Farrar model for the magnetic field
is based on the use of the NE2001 \cite{NE2001} electron densities. However as we are almost always in the small angle approximation for both $\theta$ and $\Delta$, the precise expression for the electron densities makes little difference to the results. This can be seen in the appendix, where the scale height of the thick disk component is doubled: although this is quite a substantial change, it has minimal effect on the conversion probabilities.}
 The behaviour of $P$ is very sensitive to low values of $n_{e}$, as $\tan 2 \theta$ diverges as $n_e \to 0$, so it is important that we do not allow unrealistically low values of $n_{e}$ in our calculation. The model given in \cite{Gomez} is based on line of sight integrals and gives extremely low values for $n_{e}$ at the edge of the Milky Way. For most astrophysical applications, these unphysically small values for $n_e$ have no detrimental effect, but that is not true for us.
We therefore set a minimum value for $n_{e}$ of $10^{-7} \, {\rm cm}^{-3}$, the approximate value of the intergalactic free electron density. The thick disk component is:
\begin{equation}
n_{{\rm thick}} = 1.77 \times 10^{-2} \, {\rm cm}^{-3} \times \frac{{\rm sech}^{2}\left( \frac{\rho}{15.4 \, {\rm kpc}}\right)} {{\rm sech}^{2}\left( \frac{R_{\odot}}{15.4 \, {\rm kpc}}\right)} {\rm sech}^{2}\left( \frac{z}{1.10 \, {\rm kpc}} \right) ,
\end{equation}
where $\rho$ is the cylindrical galactic radius, and
the thin disk component is:
\begin{equation}
n_{{\rm thin}} = 1.07 \times 10^{-2} \, {\rm cm}^{-3} \times \frac{{\rm sech}^{2}\left( \frac{\rho}{3.6 \, {\rm kpc}}\right)} {{\rm sech}^{2}\left( \frac{R_{\odot}}{3.6 \, {\rm kpc}}\right)} {\rm sech}^{2}\left( \frac{z}{0.04 \, {\rm kpc}} \right).
\end{equation}
For the total electron density we take
\begin{equation}
n_{e} = {\rm max} \left[n_{{\rm thick}} + n_{{\rm thin}}, 10^{-7} \, {\rm cm}^{-3} \right].
\end{equation}
A recent study by Gaensler et al \cite{Gaensler} 
suggests that the vertical scale height $h$ of the electron density may be almost double previous estimates, at $h = 1.83 \, {\rm kpc}$. We therefore also simulate the conversion probabilities with a modified electron density such that the scale height of the thick disk component is $h = 1.83 \, {\rm kpc}$.

As discussed above, the observed signal is attenuated by photoelectric absorption on gas in the galaxy. We use the scattering cross sections in \cite{Brown, MorrisonMcCammon}.
  For a 3.55 keV photon, the effective cross section per Hydrogen atom is $\sigma_{{\rm eff}} \sim 10^{-23} \, {\rm cm}^{2}$.
  For neutral hydrogen distributions, we use the density distributions for HI and H$_{2}$, $n_{HI}$ and $n_{H2}$ respectively, given in \cite{Misiriotis}:
\begin{equation}
n_{HI} =
\begin{cases}
    0.32 \, {\rm cm}^{-3} {\rm exp} \left( - \frac{\rho}{18.24 \, {\rm kpc}} - \frac{\shortmid z \shortmid} {0.52 \, {\rm kpc}} \right),& \text{if } \rho \geq 2.75 \, {\rm kpc}\\
    0,              & \text{otherwise}
\end{cases}
\end{equation}
\begin{equation}
n_{H2} = 4.06 \, {\rm cm}^{-3} {\rm exp} \left( - \frac{\rho}{2.57 \, {\rm kpc}} - \frac{\shortmid z \shortmid} {0.08 \, {\rm kpc}} \right).
\end{equation}

\section {Results and Discussion}

\subsection{Milky Way}

We simulated the propagation of ALPs created at 1 kpc, 2 kpc, ....... and 30 kpc from the Earth. Although the dark matter halo extends beyond this distance, the magnetic field at 30 kpc becomes negligible and we therefore assume that ALPs created over 30 kpc from the Earth propagate without conversion up to this point. As the magnetic field has non-trivial structure over these length scales, each of these paths must be simulated separately, rather than simply adding the probabilities. The environment experienced by the ALP, and therefore $P$, also depends strongly on where in the sky it was created. For each spherical shell, we simulated $P$ for 20000 points in the sky, equally spaced in Earth-centred spherical polar coordinates $\theta$ and $\phi$ (with $\delta \theta = \frac{\pi}{100}$ and $\delta \phi = \frac{\pi}{100}$). We used 4000 domains for each path.

The expected photon flux for each direction in the sky is:
\begin{equation}
 F = \frac{1}{\tau} \frac{1} {4 \pi} \int_{r = 0}^{\infty} \frac{\rho_{DM} \left(r, \theta, \phi \right)} {m_{DM}} P \left(r, \theta, \phi \right) dr \, {\rm sr}^{-1} ,
\end{equation}
where $P(r, \theta, \phi)$ is the ALP to photon conversion probability for an ALP starting at position $(r, \theta, \phi)$ and travelling to Earth.

We evaluate the integral as a sum for the first 30 spherical shells. For ALPs produced greater than 30  kpc from the Earth, we take $P \left( r > 30 \, {\rm kpc}, \theta, \phi \right) = P \left( 30 \, {\rm kpc}, \theta, \phi \right)$ and integrate $\rho_{DM}$ from $r = 30 \, {\rm kpc}$ to $r = \infty$ numerically.

Figure 2 shows the conversion probability for an ALP with $M = 10^{13} \, {\rm GeV}$ starting at 30 kpc from the Earth and propagating to Earth, using the original vertical scale height $h = 1.1 \, {\rm kpc}$ for the electron density. In Figure 3, we show the expected flux of the 3.55 keV line from the Milky Way halo in the ALP scenario using the NFW dark matter density and original electron density scale height with $M = \sqrt{\frac{5 \times 10^{24} \, {\rm s}}{\tau}} \times 10^{13} \, {\rm GeV}$. For comparison, Figure 4 shows the expected flux assuming the dark matter decays directly to photons, using the sterile neutrino model with the best fit neutrino decay mixing angle ${\rm sin}^{2} \left( 2 \theta \right) \simeq 7 \times 10^{-11}$ given in \cite{Bulbul}. Typical fluxes are $10^{-4} \, {\rm cm}^{-2} {\rm s}^{-1} {\rm sr}^{-1}$ for ${\rm DM} \to a \to \gamma$ and  $0.1 \, {\rm cm}^{-2} {\rm s}^{-1} {\rm sr}^{-1}$ for ${\rm DM} \to \gamma$, where the lifetime is normalised to give the same signal strength for each scenario when the dark matter decay occurs in a galaxy cluster. Equivalent plots for the Einasto dark matter distribution and $h = 1.83 \, {\rm kpc}$ are shown in the Appendix. These changes made a negligible difference to the expected flux. We also found that neglecting photoelectric absorption had only a small effect on the results, changing the expected flux by $O \left( 1 - 10 \% \right) $ with the largest affect along the direction of the galactic disk where the abundance of neutral hydrogen and heavier elements is greatest.

\begin{figure} [H]
\includegraphics[scale=0.9]{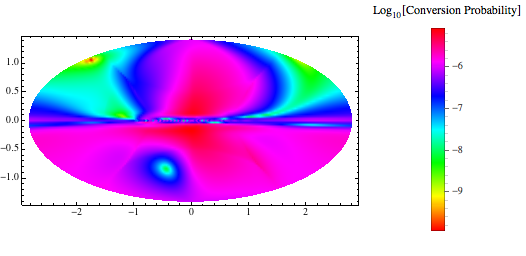}
\caption{The ALP to photon conversion probability for an ALP with $M = 10^{13} \, {\rm GeV}$ starting at 30 kpc from the Earth and propagating to Earth using a vertical scale height $h = 1.1 \, {\rm kpc}$ for the electron density. Galactic longitude increases to the right and galactic latitude increases vertically. The centre of the plot corresponds to the direction of the galactic centre.}
\end{figure}

\begin{figure} [H]
\includegraphics[scale=0.9]{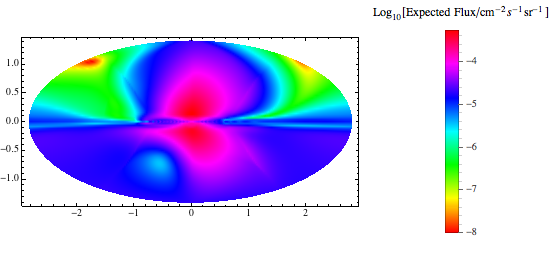}
\caption{The expected 3.55 keV line flux across the sky in the ALP scenario, using the NFW dark matter density and an electron density scale height $h = 1.1 \, {\rm kpc}$.}
\end{figure}

\begin{figure} [H]
\includegraphics[scale=0.9]{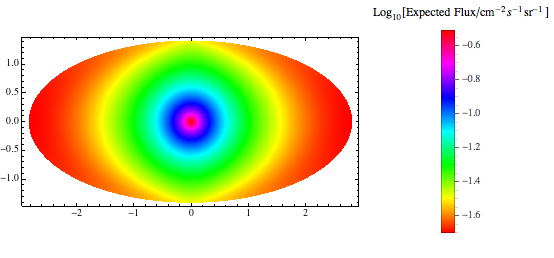}
\caption{The expected 3.55 keV line flux across the sky for direct dark matter decay to photons, using the NFW dark matter density.}
\end{figure}

We firstly note that the expected flux in the ALP scenario is almost 1000 times lower than for direct decay. This is because the ALP to photon conversion probability in the Milky Way is much lower than that in galaxy clusters, predominantly due to the Milky Way's smaller size. As discussed above, this calculation relies on approximating the ALP to photon conversion probability in all the clusters with that for the central region of Coma, and on the model used for the Milky Way's magnetic field. It might therefore be that the actual flux from the Milky Way halo is larger (or even smaller). Nonetheless, we can expect that the flux from the Milky Way in the ALP scenario would be lower than for direct decay to photons.

Let us compare the Milky Way flux we find to the projected sensitivity of the ASTRO-H mission and its Soft X-ray Spectrometer (SXS) \cite{ASTRO-H}. The maximal integrated flux we
find is $\sim 2 \ti 10^{-4} \, {\rm cm}^{-2} {\rm s}^{-1} {\rm sr}^{-1}$. The SXS will observe with an effective area of $\sim 200 \, {\rm cm}^{-2}$ and with a field of view of
$\sim 3^{'} \ti 3^{'}$. The resulting count rate on the SXS from $a \to \gamma$ is then at most $\sim 4 \ti 10^{-8} \, {\rm s}^{-1}$, rendering detection entirely impossible.
It is here we see the significance of the low $a \to \gamma$ conversion probabilities for the Milky Way compared to galaxy clusters. Compared to clusters,
the conversion probability in the Milky Way halo is reduced
by a factor of $10^3$, making the signal entirely unobservable. This is in contrast to the case of direct dark matter decay to photons, where as explained in
\cite{Boyarsky} a failure to detect the 3.55 keV line from the Milky Way halo would rule out the model. Indeed, a recent analysis of {\it Chandra} X-ray observations \cite{Riemer-Sorensen} failed to detect the 3.55 keV line in the Milky Way. This result is in strong tension with models in which the dark matter decays directly into a photon, but is consistent with the $DM \to a \to \gamma$ scenario.

\subsection{Andromeda (M31)}

Given an observation of a 3.55 keV photon line from M31 \cite{Boyarsky}, it is also important to consider the magnitude of ALP-photon conversion in M31, as M31 is in many ways a similar galaxy to the Milky Way - although twice the size with a diameter 
of around 70 kpc compared to 30 kpc for the Milky Way, it is also a spiral galaxy.
Naively, one would therefore expect a similar suppression of the ALP-photon conversion probailities in M31. However this appears not to be true, as radio studies \cite{Han98, Fletcher, Egorov} show that the regular magnetic field in M31 is both significantly larger and significantly more coherent than in the Milky Way.

Let us summarise the results of \cite{Han98, Egorov, Fletcher}. \cite{Fletcher} found that between 6 and 14 kpc from the centre of M31, the M31 magnetic field is a coherent axisymmetric spiral, with a regular magnetic field strength $B_{reg} \sim 5 \, \mu {\rm G}$ that showed minimal radial variation over this range. In contrast to the Milky Way, there is also no sign of large-scale field reversals among the spirals. They also found that the vertical scale height of this field was at least 1 kpc. \cite{Han98} found that the regular magnetic field of M31 probably has a similar structure between radii of 5 and 25 kpc. The random magnetic field of M31 is a similar size to the regular field, $B_{random} \sim 5 \, \mu G$. As discussed in \cite{HealdBeck2013}, the fact that $B_{regular} \sim B_{random}$ in M31 is quite an unusual property, as generally $B_{regular} \sim \frac{B_{random}}{3}$ (as holds in the
Milky Way).

The central magnetic field in M31 is also rather large. For example, based on measurements in the central kiloparsec the authors of \cite{Egorov} estimate a central field $B \simeq 50 \, \mu {\rm G}$.
Another key point is that the plane of the M31 disk is at an inclination angle of $77.5^{\circ}$ (where $90^{\circ}$ corresponds to an entirely edge on spiral galaxy). This is close to edge on, and implies ALPs originating from dark matter decay in M31 pass through a large coherent transverse magnetic field on their way to Earth.
These are optimal conditions for $a \to \gamma$ conversion. This would \emph{not} have been true if M31 has an inclination angle close to zero.

Given the above, we use the single domain small angle approximation to obtain a rough estimate of the conversion probability for a 3.55 keV ALP created at the centre of M31 and propagating to Earth.
We take parameters appropriate for propagation directly along the disk. This approximation is an overestimate based on a fall-off in the magnetic field both out of the plane of the disk and with radius, and an underestimate based on neglecting the strong central magnetic field and any halo or out-of-plane components of the M31 magnetic field. Taking $B_{\perp} \sim 5 \, \mu {\rm G}$, and $L \sim 20 \, {\rm kpc}$, we estimate
\be
P_{a \to \gamma, M31} \sim 2.3 \ti 10^{-4} \left( \frac{10^{13} \, \hbox{GeV}}{M} \right)^2,
\ee
which is two orders of magnitude higher than typical conversion probabilities for the Milky Way. The difference arises from the greater magnitude and coherence length of the M31 magnetic field compared to that of the Milky Way, coupled to the close to edge on nature of M31. In fact, the above conversion probability is only smaller by a factor of four than that found for a $1 \, \hbox{Mpc}$ path through the central region of the Coma cluster. This shows that in this scenario the observed signal strength from M31 can be comparable
to that from clusters, consistent with the results of \cite{Boyarsky}.

We also note that the spiral regular magnetic field in M31 implies that for observations offset from the centre we expect a rapid falloff in the signal. As we move away from the centre, the field lines of the regular spirals will become parallel to the observational line of sight rather than transverse to the line of sight. As ALP-photon conversion only occurs for transverse magnetic fields, this will rapidly reduce the signal strength. ALP-photon conversion will still occur due to the random component of the M31 magnetic field. However, as the coherence scale $L$ of the random field is expected to be similar to the Milky Way at $\sim 100 {\rm pc}$, and the ALP-photon conversion probability scales as $P(a \to \gamma) \sim L^2$,
we expect that conversion due to the random field will be negligible compared to that due to the regular field.
This is consistent with the absence, albeit at low statistics, of an observed off-centre M31 line in \cite{Boyarsky}.

We note that  - within the ${\rm DM} \to a \to \gamma$ scenario - the above points make M31 an unusually favourable galaxy for observing a 3.55 keV line. For general galaxies in this scenario the signal strength of the 3.55 keV line would be much lower than for galaxy clusters, and the fact that for M31 these can be comparable is rather uncommon.

\section {Conclusions}

In this paper we have studied the expected signal from the process ${\rm DM} \to a \to \gamma$ in the Milky Way halo, assuming that the same process is responsible for the 3.55 keV line observed in both galaxy clusters and M31. We have also studied the same process in M31.

In doing so we have used the recent model of \cite{Farrar} for the magnetic field of the Milky Way. We note this model artificially excludes the magnetic field within 1 kpc of the galactic centre, which may be significantly higher than the bulk of the Milky Way. Our results therefore apply clearly to the Milky Way halo as a whole, but only apply for sightlines passing close to the galactic centre on the (strong) assumption that the magnetic field in that region is not significantly stronger than in the Milky Way as a whole.

We can therefore make the following predictions for a 3.55 keV photon line from the Milky Way halo in the case that the dark matter decays to an ALP:

\begin {enumerate}

\item In this scenario, the flux from the Milky Way halo will be significantly lower than for the case of direct dark matter decay to photons. This arises as the ALP to photon conversion probability in the Milky Way is much lower than in galaxy clusters. This is due to both the relatively small magnetic field and relatively small coherence length in the Milky Way.

\item The flux from the Milky Way halo will be unobservable with ASTRO-H, unless the magnetic fields in both galaxy clusters and M31 have been significantly overestimated, or the Milky Way magnetic field has been significantly underestimated.

\item Although M31 is in some ways similar to the Milky Way, the conversion probabilities for ${\rm DM} \to a \to \gamma$ for M31 are larger by approximately two orders of magnitude. This is because M31 is close to edge on to us, with a large regular magnetic field coherent over a large distance. In fact the $a \to \gamma$ conversion probabilities for ALPs travelling through M31 to us are comparable to those for clusters.

\item A non-observation of the 3.55 keV line from the Milky Way with ASTRO-H will not rule out a dark matter origin of the signal.

\end {enumerate}

\section*{Acknowledgments}

JC is funded by a Royal Society University Research Fellowship and by the ERC Starting Grant `Supersymmetry Breaking in String Theory'.
FD is funded by an STFC studentship. We thank Pedro Alvarez, Stephen Angus, Felix Kahlhoefer, David Marsh, Andrew Powell and Markus Rummel for discussions.

\appendix
\section{The Effect of Varying the Dark Matter and Electron Density Distributions}
As described in Section 3, there are a number of possible choices for the distributions of dark matter and free electrons in the galaxy. In Figures 2 - 4, we used the NFW dark matter distribution and a vertical scale height $h = 1.1 \, {\rm kpc}$ for the electron density. Here we show equivalent figures using the Einasto dark matter distribution and $h = 1.83 \, {\rm kpc}$. Comparing these to Figures 2 - 4 we see that these changes make a negligible difference to the expected flux.

\begin{figure} [H]
\includegraphics[scale=0.9]{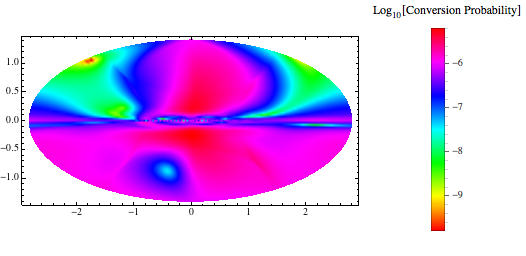}
\caption{The ALP to photon conversion probability for an ALP with $M = 10^{13} \, {\rm GeV}$ starting at 30 kpc from the Earth and propagating to Earth using a vertical scale height $h = 1.83 \, {\rm kpc}$ for the electron density. Galactic longitude increases to the right and galactic latitude increases vertically. The centre of the plot corresponds to the direction of the galactic centre.}
\end{figure}

\begin{figure} [H]
\includegraphics[scale=0.9]{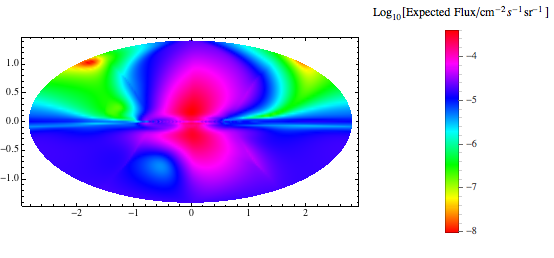}
\caption{The expected 3.55 keV line flux across the sky in the ALP scenario, using the Einasto dark matter density and an electron density scale height $h = 1.83 \, {\rm kpc}$.}
\end{figure}

\begin{figure} [H]
\includegraphics[scale=0.9]{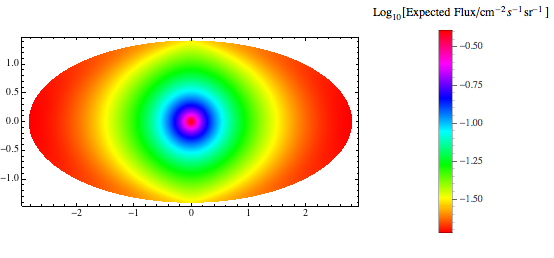}
\caption{The expected 3.55 keV line flux across the sky for direct dark matter decay to photons, using the Einasto dark matter density.}
\end{figure}

\bibliography{refs}
\bibliographystyle{JHEP}

\end{document}